\documentclass[preprint,12pt]{elsarticle}

\usepackage{graphicx}
\usepackage{epsfig}

\usepackage{amssymb}

\newcommand{\cp}{\chi^{(+)}}
\newcommand{\cm}{\chi^{(-)*}}

\newcommand{\vv}{V_{bn}({\bf r}_1)}
\newcommand{\ri}{{\bf r}_i}
\newcommand{\ro}{{\bf r}_1}
\newcommand{\ak}{{\bf k}_a}
\newcommand{\bq}{{\bf k}_b}

\newcommand{\rn}{{\bf r}_n}
\newcommand{\nq}{{\bf k}_n}
\newcommand{\we}{\Psi^{(+)}_a(\xi_a,{\bf r}_1,{\bf r}_i)}

\begin{document}

\begin{frontmatter}



\title{Coulomb breakup of $^{37}$Mg and its ground state structure}



\author{Shubhchintak}
\ead{shub1dph@iitr.ac.in}
\author{Neelam}
\ead{nph10dph@iitr.ac.in}
\author{R. Chatterjee}
\ead{rcfphfph@iitr.ac.in}
\address{Department of Physics, Indian Institute of Technology - Roorkee, 247667, 
India}
\author{R. Shyam}
\ead{radhey.shyam@saha.ac.in}
\address{Theory Group, Saha Institute of Nuclear Physics, 1/AF Bidhannagar,
Kolkata 700064, India, and Department of Physics, Indian Institute of Technology,
Roorkee-247667, India}
\author{K. Tsushima}
\ead{kazuo.tsushima@gmail.com}
\address{International Institute of Physics, Federal University of Rio Grande
do Norte Av. Odilon Gomes de Lima, 1722 Capim Macio, Natal, RN 59078-400, Brazil}

\begin{abstract}

We calculate Coulomb breakup of the neutron rich nucleus $^{37}$Mg on a Pb 
target at the beam energy of 244 MeV/nucleon within the framework of a finite 
range distorted wave Born approximation theory that is extended to include the 
effects of projectile deformation. In this theory, the breakup amplitude involves 
the full wave function of the projectile ground state. Calculations have been 
carried out for the total one-neutron removal cross section $(\sigma_{-1n})$, 
the neutron-core relative energy spectrum, the parallel momentum distribution 
of the core fragment, the valence neutron angular, and energy-angular distributions.
The calculated $\sigma_{-1n}$ has been compared with the recently measured 
data to put constraints on the spin parity, and the one-neutron separation energy 
($S_n$) of the $^{37}$Mg ground state ($^{37}$Mg$_{gs}$). The dependence of 
$\sigma_{-1n}$ on the deformation of this state has also been investigated. While  
a spin parity assignment of $7/2^-$ for the $^{37}$Mg$_{gs}$ is ruled out by our 
study, neither of the $3/2^-$ and $1/2^+$ assignments can be clearly excluded. 
Using the spectroscopic factor of one for both the $3/2^-$ and $1/2^+$ 
configurations and ignoring the projectile deformation effects, the $S_n$ 
values of $0.35 \pm 0.06$ MeV and $0.50 \pm 0.07$ MeV, respectively, are extracted 
for the two configurations. However, the extracted $S_n$ is strongly dependent
on the spectroscopic factor and the deformation effects of the respective 
configuration. The narrow parallel momentum distribution of the core fragment 
and the strong forward peaking of the valence neutron angular distribution suggest 
a one-neutron halo configuration in either of the $2p_{3/2}$ and $2s_{1/2}$ 
configurations of the $^{37}$Mg ground state. 
\end{abstract}


\end{frontmatter}


\section{Introduction}
\label{1}

With the advances made in the technology of producing nuclear species with
relatively large neutron ($N$) to proton ($Z$) number ratios, it is now 
possible to extensively study nuclei near the neutron-drip line with $Z > 8$. 
During the last three decades measurements performed on mass, radius and 
spectroscopy of such nuclei have shown that they have structures that are at 
variance with those of their ``near the line of stability" counterparts (see, e.g.,
\cite{thi75,hub78,det79,gui84,bau89,mot95,pri00,iwa01,yan03,chu05,ney05,yor07},
and~\cite{gad07,him08,sor08,doo09,kan10,wim10,hin11,gad11,doo13}). With the advent
of new generation of radioactive ion beam facilities, it has now become possible
not only to produce medium mass neutron rich nuclei in the vicinity of the magic 
numbers but also employ them as projectiles to initiate reactions (e.g., breakup) 
on nuclear targets~\cite{nak09,kob14}. This provides excellent opportunity to 
perform quantitative study of the single-particle structure and the shell evolution 
in this region.

The notion of ``magic" numbers is one of the most fundamental concepts in nuclear
structure physics~\cite{hax49,may49}. If large gaps occur between groups of
single-particle orbits that are completely filled with nucleons (neutrons or 
protons), then these nucleon numbers are called ``magic". The seven most established 
magic numbers are 2, 8, 20, 28, 50, 82, and 126. However, in several nuclei near 
the neutron-drip line, modifications to this shell structure have been observed
\cite{sor08}. In these cases the magic numbers evolve as a function of the neutron
number - old magic numbers may disappear while new ones emerge and conventional
shell gaps may break down. The region where abrupt onset of changes in the magic
numbers appears, is called island of inversion \cite{war90}. For example, rapid
changes in nuclear structure and vanishing of the $N = 8$, and $20$ shell gaps have
been seen in neutron rich nuclei $^{12}$Be~\cite{iwa00}, and $^{32}$Mg~\cite{mot95}
and $^{30,32}$Ne~\cite{doo09}, respectively. Examples of $N = 28$ shell quenching
have been observed in $^{36,38}$Mg~\cite{doo13} and $^{42}$Si~\cite{bas07}. It is
suggested in Ref.~\cite{war90} that island of inversion near $N = 20, 28$ comes
about because of the fact that the $\nu(sd)^{-2}(fp)^2$ intruder configurations
(here $\nu$ represents a relative neutron state), in which two neutrons from the
$sd$ shell are excited to the $fp$ shell, become so low in energy that they form
the ground states for $Z = 10 - 12$ and $N = 20 - 22$ nuclei. This suggestion was
confirmed subsequently by mass measurements of the neutron rich isotopes of Ne, Na
and Mg nuclei~\cite{orr91}. Recently, this behavior has been shown to be a general
phenomena that should occur for most standard shell closures far from the line of
stability, and the mechanism behind this is found to be related to the importance
of the nucleon-nucleon tensor interaction~\cite{ots05}. It is obvious that due to
the intruder states, the single particle structure of the ground states of nuclei
lying within island of inversion will not be the same as that emerging from the
usual filling of the shell model states.

The mixing of neutron n-particle-n-hole ($np-nh$) intruder configurations of
$\nu(sd)^{-n}(fp)^n$ character to the ground state, causes large deformation to
nuclei in island of inversion near $N = 20, 28$, which is confirmed by the measured
low excitation energies and $B(E2)$ values of the first excited states~(see, eg.,
Refs.~\cite{det79,mot95,iwa01,gad07,doo09,yon01}).  It has been emphasized
\cite{war90,cam75,pat91,pov94} that the deformation may also account for the
enhanced binding energies manifested in some of the known nuclei in this region.
The collective properties of neutron rich nuclei near $N =20 $ region are rather
well described by state-of-the-art Monte-Carlo shell model calculations that allow
for unrestricted mixing of neutron particle-hole configurations across the shell
gap~\cite{uts99,ots01}.  In Ref.~\cite{ham07,ham12}, nuclei in the neighborhood of
neutron-drip line have been systematically investigated in a model where
one-particle motion is described within spherical as well as deformed potentials.
It has been concluded in this work that nuclei in the region of $N = 20 - 28$ are
most likely to be deformed.

The root mean square radius (RMS) of a deformed  nucleus becomes effectively
larger than that of a spherical one. This enhances the total reaction cross
section ($\sigma_R$) that depends on RMS radii of the projectile and the target
nuclei. Large interaction cross sections (which are almost the same as the
$\sigma_R$) have been measured for $^{29-32}$Ne~\cite{tak10,tak12} and
$^{24-38}$Mg~\cite{tak14} nuclei. This has led to the conclusion that the isotopes
$^{29-32}$Ne have strong deformation~\cite{min12,sum12,hor12}. For the Mg case,
these studies~\cite{hor12,wat14} suggest that while $^{27}$Mg and $^{30}$Mg are
spherical, $^{25,29,33-38}$Mg are more likely to be deformed.

The discovery of halo structure in some of the drip line nuclei is another
important progress made in the studies of nuclei with large $N$ to $Z$ ratio near
the limits of nuclear stability~\cite{tan85,han87,ber93,pra99,jon04,fre12,tan13}.
As the neutron-drip line is approached nuclei experience weakening of the neutron
binding energy, which leads to some special effects. The sudden rise of interaction
cross sections with increasing $N$ in some of these nuclei can be attributed to the
extended density distribution(s) of the valence neutron(s). This decoupling of the
valence neutron(s) from the tightly bound core and the extension of the
corresponding wave function to much larger radii have been referred to as neutron
halo. This phenomena has been seen earlier in lighter nuclei like $^{11}$Li,
$^{11}$Be, $^{19}$C~\cite{tan85}.

In recent years, there has been considerable interest in finding out if halo
configurations also exist in nuclei lying near the neutron-drip line in the
vicinity of island of inversion~\cite{rot09,cha13}. $^{31}$Ne with $Z=10$ and
$N=21$, is a promising candidate to have a one-neutron halo configuration because
the one-neutron separation energy of this nucleus is quite small ($0.29 \pm 
1.64$ MeV~\cite{jur07}, or $0.06 \pm 0.41$ MeV~\cite{gau12}). Indeed, such a 
structure has been suggested for this isotope by Coulomb breakup studies 
\cite{nak09}. This has been further supported by measurements~\cite{tak10} of 
the interaction cross sections for Ne isotopes incident on a $^{12}$C target at
the beam energy of 240 MeV/nucleon, where it was found that for $^{31}$Ne the
interaction cross section was much larger than that of any other Ne isotope.
Recently, measurements of $\sigma_R$ for $^{24-38}$Mg isotopes on $^{12}$C 
target at the beam energy of 240 MeV/nucleon have been reported in 
Ref.~\cite{tak14}. From a similar reasoning, it was suggested in this study that 
$^{37}$Mg ($Z=12, N=25$) that lies in $N=20-28$ island of inversion, is also a
candidate for having a one-neutron halo structure. This was reinforced by
measurements of Coulomb breakup of $^{37}$Mg on C and Pb targets at the beam
energy of 244 MeV/nucleon in Ref.~\cite{kob14}.

The observation of the halo phenomena in the heavier nuclei lying in island of
inversion, signals major changes in the shell evolution in these nuclei as
compared to that seen in the spherical ones. One of the conditions for the halo
formation is that the loosely bound neutron in the nucleus occupies a low orbital
angular momentum state ($\ell = 0 $ or 1) in order to reduce the centrifugal
barrier effects that prevent it from spreading out~\cite{rot09}. In fact, in
well established cases of light one-neutron halo nuclei like $^{11}$Be and
$^{19}$C, the ground states have predominant $s$-wave neutron plus core
configurations~\cite{pra99,nak94,nun96,aum00,pal03}. According to the 
conventional shell model evolution, one expects to see the domination of the 
1$f_{7/2}$ orbit in nuclei in the vicinity of $N=20 - 28$. This would not favor 
the halo formation because a larger centrifugal barrier would prevent the $l=3$ 
neutrons to extend too far out in the space. Therefore, a significant contribution 
from the $s$- or $p$-wave orbits has to be there in the ground state structure of 
these nuclei to minimize the centrifugal barrier. Thus, the existence of halo 
structure would imply a significant modification of the shell structure that 
involves considerable mixing of the intruder states like $2p_{3/2}$ or $2s_{1/2}$ 
into the ground states of these nuclei, which also leads to the appreciable 
deformation of these states. Therefore, the halo formation in heavier nuclei in 
island of inversion region has strong correlation with the shell evolution and the 
presence of deformation.

Coulomb breakup reaction, in which the valence neutron is removed from the fast
projectile in Coulomb fields of heavy target nuclei, provides a convenient tool
to investigate the halo structure in the neutron-drip line nuclei~(see, e.g., the
review~\cite{bau03}). It places constraints on their electric dipole response
\cite{nak94,nak99,bau03}. A class of theories of this reaction (e.g., the post
form finite range distorted wave Born approximation (FRDWBA) theory
\cite{pra99,cha00}) requires realistic wave functions to describe the relative
motion between the valence neutron and the core in the ground state of the
projectile. Thus by comparing the calculations of the cross sections with the
measured data, one can directly probe the ground state structure of the projectile.
Narrow widths of the parallel momentum distributions of the core fragments provide
a robust signature of the presence of halo structure in the projectile nuclei as
they imply a larger spatial spread of the fragments in their ground states.

Recently~\cite{shu14}, the FRDWBA theory of Coulomb breakup reactions has been
extended to include the deformation of the projectile by using a deformed
Woods-Saxon potential to describe the valence neutron-core relative motion.
Hence, this provides a microscopic theoretical tool to study the Coulomb breakup 
of neutron-drip line nuclei lying in island of inversion in the vicinity of $N = 
20 - 28$ and to investigate the correlation between halo formation and the shell
evolution and deformation. In the first application of this theory, the Coulomb
breakup of $^{31}$Ne on a Pb target at the beam energy of 234 MeV/nucleon was
investigated~\cite{shu14}. Comparison of calculated and the measured $\sigma_{-1n}$ 
(of Refs.~\cite{nak09,nak14}) suggested that the ground state of $^{31}$Ne is 
most likely to have a $^{30}$Ne $\otimes$ $2p_{3/2}\nu$ configuration. The value 
of the one-neutron separation energy $S_n$ is found to be correlated to the 
quadrupole deformation parameter ($\beta_2$). For $\beta_2$ between $0.0-0.5$,
$S_n$ varies between $0.24-0.58$ MeV. The calculated full width at half maximum
(FWHM) of the parallel momentum distribution of $^{30}$Ne fragment is closer to 
that of the core fragment seen in the breakup reaction of the established halo
nucleus $^{11}$Be. This strongly suggests that the ground state of $^{31}$Ne has 
a one-neutron halo structure in the $2p_{3/2}$ state.

$^{37}$Mg is the most neutron-rich bound odd-mass Mg isotope. However, the
experimental information about its mass and the ground state spin parity is not
available, even though mass systematics suggest that it is a very weakly bound
system with $S_n$ in the range of 0.16 $\pm$ 0.68 MeV~\cite{wan12}. Therefore,
it is another promising candidate for having a one-neutron halo structure in 
island of inversion near $N = 20-28$. However, a $1f_{7/2}$ configuration for 
its ground state that would result in the conventional spherical shell model, will 
suppress the halo formation due to the high centrifugal barrier. Hence, a 
significant modification of its spherical shell structure with introduction of 
the intruder configurations having $s$- and $p$-wave states, is necessary for this 
nucleus to have a halo like structure. This also implies that its ground state 
should be deformed.

The aim of this paper is to investigate the one-neutron removal cross section of
$^{37}$Mg on a Pb target at the beam energy of 244 MeV/nucleon within the FRDWBA
theory of Coulomb breakup reactions. By comparing the $\sigma_{-1n}$ calculated
within this theory with the corresponding experimental data we would like to
extract most plausible spin parity for the ground state of this nucleus. We
attempt to put constraints on the large uncertainty in its $S_n$ value. We
investigate the effect of the ground state deformation of this nucleus on the
values of $\sigma_{-1n}$ deduced in our analysis. Furthermore, we make predictions
for the observables such as relative energy spectra of the valence neutron-core
fragments, parallel momentum distribution of the core fragment, and the angular
distribution of the valence neutron as a function of deformation. Our study is
expected to provide more understanding about the evolution of the shell structure 
in island of inversion from $N=$ 20 to 28, about which some conflicting results 
have been reported recently~\cite{gad07,doo13}. Furthermore, our study is expected 
to quantify the presence of a neutron halo structure in $^{37}$Mg and provide
information about its correlation to the ground state deformation of this nucleus.

The study of Coulomb breakup of $^{37}$Mg is also of interest in astrophysics 
because it provides an indirect way to determine the rate of the radiative neutron 
capture reaction $^{36}$Mg($n,\gamma$)$^{37}$Mg, which is of importance in the 
study of the $r$-process nucleosynthesis in supernovae~\cite{ter01}.

In the next section, we present our formalism where we recall some important 
aspects of the FRDWBA theory of breakup reactions and its extension to include 
the deformation of the projectile ground state. The results of our calculations 
are presented in section 3, where we discuss the one-neutron removal cross section, 
the relative energy spectra of the fragments, the parallel momentum distribution 
of the core fragment, the angular and energy distribution of valence neutron as a 
function of the projectile deformation. The summary and conclusions of our study 
are presented in section 4.


\section{Formalism}

The breakup reaction of a projectile $a$ into the core fragment $b$ and the valence 
neutron $n$, in the Coulomb field of a target $t$ can be represented as $ a + t 
\rightarrow b + n + t $. We assume that target nucleus remains in its ground state 
during the breakup process. Thus this is also known as the elastic breakup 
reaction. The chosen coordinate system is shown in Fig.~1.
\begin{figure}[here]
\begin{center}
\mbox{\epsfig{file=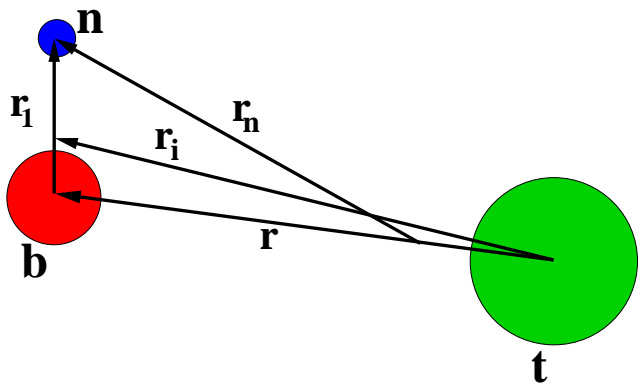,height=5.0cm}}
\end{center}
\caption{
The three-body coordinate system. The charged core fragment, the valence neutron 
and the target nucleus are denoted by $b$, $n$ and $t$, respectively.
}
\label{fig:figa)}
\end{figure}
\noindent
The position vectors satisfy the following relations:
\begin{eqnarray}
{\bf r} &=& \ri - \alpha\ro,~~ \alpha = {m_n\over {m_n+m_b}} \,
,   \\
\rn &=& \gamma\ro +\delta\ri, ~~ \delta = {m_t\over {m_b+m_t}},
~~  \gamma = (1 - \alpha\delta) \, .
\end{eqnarray}

The starting point of the FRDWBA theory of Coulomb breakup is the 
post-form $T$-matrix of the reaction given by
\begin{eqnarray}
T = \int  d\xi d\ro d\ri \cm_b(\bq,{\bf r})\Phi^*_b(\xi_b)
\cm_n(\nq,{\bf \rn})\Phi^*_n(\xi_n)\vv \we.
\end{eqnarray}
The functions $\chi$ are the distorted waves for the relative motions of $b$ 
and $n$ with respect to $t$ and the center of mass (c.m.)\ of the $b+t$ system, 
respectively.  The functions $\Phi$ are the internal state wave functions of 
the concerned particles dependent on the internal coordinates $\xi$. The 
function $\we$ is the exact three-body scattering wave function of the 
projectile with a wave vector $\ak$ satisfying outgoing boundary conditions.  
The vectors $\bq$ and $\nq$ are the Jacobi wave vectors of $b$ and $n$, 
respectively, in the final channel of the reaction. The function $\vv$ 
represents the interaction between $b$ and $n$.  As we concentrate only on 
Coulomb breakup, the function $\chi^{(-)}_b({\bq},{\bf r})$ is taken as the 
Coulomb distorted wave (for a point Coulomb interaction between the charged
core $b$ and the target) satisfying incoming wave boundary conditions, and
the function $\chi^{(-)}_n({\nq},{\rn})$ describing the relative motion of the 
neutron with respect to the target, is just a plane wave. It may be noted that
within this approach the fragment-target interactions are treated to all orders. 

In the distorted wave Born approximation (DWBA), we write
\begin{eqnarray}
\we = \Phi_a(\xi_a,\ro)\cp_a(\ak,\ri),
\end{eqnarray}
The assumption inherent in Eq.~(4) is that the breakup channels are very weakly 
coupled and hence this coupling needs to be treated only in the first order. 
We expect this approximation to be valid for those cases where there are no 
resonances in the $n + b$ continuum. Most of the neutron halo systems come in
this category. For those cases where higher order effects of fragment-fragment 
interaction are non-negligible, the applicability of this method would be limited.  
Ideally, a  rigorous description of the breakup process of all types of projectiles 
would require the use of Faddeev type of three-body methods that include Coulomb 
potentials in the fragment-target and fragment-fragment (if required) interactions. 
A few such calculations have become available recently although they are confined 
to the breakup reactions on a proton target~\cite{del09,cre09,cra10,upa12}.

In Eq.~(4) the dependence of $\Phi_a$ on ${\bf r}_1$ describes the relative motion 
of the fragments $b$ and $n$ in the ground state of the projectile. The function 
$\cp_a(\ak,\ri)$ is the Coulomb distorted scattering wave describing the relative 
motion of the c.m.\ of the projectile with respect to the target, satisfying 
outgoing wave boundary conditions.  

The integration over the internal coordinates $\xi$ in the $T$-matrix gives
\begin{eqnarray}
\int d\xi\Phi^*_b(\xi_b)\Phi^*_n(\xi_n)\Phi_a(\xi_a,\ro)
& = & \sum_{\ell mj\mu} \langle \ell mj_n\mu_n|j\mu\rangle
 \langle j_b\mu_bj\mu|j_a\mu_a\rangle i^\ell \Phi_a(\ro),
\end{eqnarray}
The wave function $\Phi_a(\ro)$ can be expressed in terms of its radial and
angular parts assuming a particular partition, in which the relative motion 
between $n$ and $b$ has an orbital angular momentum $\ell$ as
\begin{eqnarray}
\Phi_a(\ro) & = & \sqrt{C^2S}\, u_{\ell}(r_1) Y_{\ell}^m({\hat{\bf r}}_1),
\end{eqnarray}
where $C^2S$ is the spectroscopic factor for the given partition. In Eq.~(5),  
$\ell$ is coupled to the spin of $n$ and the resultant channel spin $j$ is 
coupled to the spin $j_b$ of the core $b$ to yield the spin of $a$ ($j_a$).
The partition represented by Eq.~(6) will be retained even if the potential 
$\vv$ is deformed.

The $T$-matrix can now be written as
\begin{eqnarray}
T & = & \sum_{\ell mj\mu} \langle \ell mj_n\mu_n|j\mu\rangle
\langle j_b\mu_bj\mu|j_a\mu_a\rangle i^\ell
\hat{\ell}\beta_{\ell m}(\bq,\nq;\ak),
\end{eqnarray}
where
\begin{eqnarray}
\hat{\ell}\beta_{\ell m}(\bq,\nq;\ak) &=&\int d\ro d\ri\cm_b(\bq,{\bf r})
e^{-i\nq.\rn} \nonumber\\
&\times&\vv u_\ell (r_1) Y_{\ell}^m({\hat{\bf r}}_1)\cp_a(\ak,\ri).
\end{eqnarray}
with $\beta_{\ell m}$ being the reduced $T$-matrix and with ${\hat \ell} \equiv 
\sqrt{2\ell + 1}$.

Eq.~(8) involves a six dimensional integral which makes the computation of 
$\beta_{\ell m}$ quite complicated. The problem gets further acute because the 
integrand involves a product of three scattering waves that exhibit oscillatory 
behavior asymptotically. In the past calculations have been simplified by using 
approximate methods, such as the zero-range approximation (see e.g.,\ 
\cite{sat83,aus70,gle83}) or the Baur-Trautmann approximation~\cite{bau72,bau84} 
that led to the factorization of the T-matrix into two terms; each involving 
three-dimensional integrals (we refer to~\cite{shy01} for a detailed discussion).  
However, these methods are not valid for breakup reactions at higher beam energies 
and for heavier projectiles that can have non-$s$-wave ground states. 

In our FRDWBA theory we use a method that was proposed in Ref.~\cite{bra74} for 
describing the heavy ion induced transfer reactions, and was adopted in Ref.
\cite{shy85} for describing the breakup reactions of heavy projectiles. This was 
shown~\cite{cha00,shy01} to be well suited for calculating the Coulomb breakup of 
halo nuclei. In this method, the Coulomb distorted wave of particle $b$ in the 
final channel is written as~\cite{cha00}
\begin{eqnarray}
\chi^{(-)}_b(\bq,{\bf r}) & = & e^{-i\alpha{\bf K}.\ro}
                           \chi^{(-)}_b(\bq,\ri).
\end{eqnarray}
Eq.~(9) represents an exact Taylor series expansion about ${\bf r}_i$ if 
$ {\bf K} = -i\nabla_{{\bf r}_i}$ is treated exactly. However, instead of doing 
this we employ a local momentum approximation (LMA) where the magnitude of 
momentum ${\bf K}$ is taken to be
\begin{eqnarray}
K(\mathcal{R}) & = &{\sqrt {{2m\over \hbar^2}(E - V(\mathcal{R}))}}.
\end{eqnarray}
Here $m$ is the reduced mass of the $b-t$ system, $E$ is the energy of particle 
$b$ relative to the target in the c.m.\ system and $V(\mathcal{R})$ is the  Coulomb 
potential between $b$ and the target separated by $\mathcal{R}$. Thus, the magnitude 
of the momentum of ${{\bf K}}$ is evaluated at some separation $\mathcal{R}$, which 
is held fixed for all the values of ${r}$. The value of $\mathcal{R}$ was taken to 
be equal to 10 fm. For reactions under investigation in this paper, the magnitude of 
${\bf K}$ remains constant for distances larger than 10 fm. Due to the peripheral 
nature of the breakup reaction, the region $\mathcal{R} \gtrsim 10$ fm contributes 
maximum to the cross section.  Furthermore, the results of the calculations for 
these reactions, at the beam energies under investigation, are almost independent 
of the choice of the direction of momentum ${\bf K}$ \cite{cha00}.  Therefore, we 
have taken the directions of ${{\bf K}}$ and ${\bf k_b}$ to be the same in all the 
calculations presented in this paper.

Substituting Eq. (9) into Eq.~(8), the reduced amplitude is obtained in a 
factorized form as
\begin{eqnarray}
\hat{\ell}\beta_{\ell m} &=& \int d{\bf r}_i e^{-i\delta {\bf q}_n.{\bf r}_i}
\chi^{(-)*}_b({\bf q}_b,{\bf r}_i)\chi^{(+)}_a({\bf q}_a,{\bf r}_i)
\nonumber \\
& \times &
\int d{\bf r}_1 e^{-i{\bf Q}.{\bf r}_1} \vv u_\ell (r_1) Y_{\ell}^m({\hat{\bf r}}_1),
\end{eqnarray}
where, ${\bf Q} = \gamma {\bf q}_n-\alpha{\bf K}$. The first integral in Eq. (11), 
is the dynamics part in the Coulomb breakup and is expressed analytically in terms 
of the Bremsstrahlung integral \cite{nor54}. The second integral in Eq.~(11) 
contains the projectile structure information.

We now introduce deformation in potential $\vv$ in Eq.~(11). Following~\cite{ham04}, 
we write the axially symmetric quadrupole-deformed Woods-Saxon potential (without 
taking the spin-orbit term) as
\begin{eqnarray}
\vv = V_{ws}(r_1) - \beta_2 k(r_1) Y^{0}_{2}(\hat {\bf r}_1). 
\label{a4.3}
\end{eqnarray}
We take the Woods-Saxon form for the potential $V_{ws}(r_1)$, and write, 
$V_{ws}(r_1) = V_{ws}^0 f(r_1)$, where $V_{ws}^0$ is the depth of the potential 
and $f(r_1)$ describes its shape. $f(r_1)$ and $k(r_1)$ are defined as 
\begin{eqnarray}
f(r_1) = \frac{1}{1+exp(\frac{r_1-R}{a})}, \hspace{0.2in} k(r_1) = RV_{ws}^0 
\frac{df(r_1)}{dr_1}, \nonumber
\end{eqnarray}
with radius $R = r_0A^{1/3}$ where $r_0$ and $a$ are the radius and diffuseness 
parameters, respectively. $\beta_2$ is the quadrupole deformation parameter.
In Eq.~(12), we have included only the lowest-order term in the deformation 
parameter of the deformed Woods-Saxon potential ( see, e.g., Ref.~\cite{fal90}). 
This is an approximation. However, this should be sufficient for our purpose of 
illustrating the role of projectile deformation effects on the breakup cross 
sections.

The radial wavefunction corresponding to the potential $\vv$ should be obtained 
by solving the coupled equation
\begin{eqnarray}
\left\{ \frac{d^2}{dr_1^2} -\frac{\ell(\ell+1)}{r_1^2} +\frac{2\mu}{\hbar^2}
[E - V_{ws}(r_1)]\right\} u_{\ell m}(r_1)\nonumber \\ = 
\frac{2\mu}{\hbar^2}\sum_{\ell^\prime} \langle Y_{\ell}^m(\hat {\bf r}_1)|- \beta_2 k(r_1) 
Y^{0}_{2}(\hat {\bf r}_1)|Y_{\ell^\prime}^m(\hat {\bf r}_1)\rangle  u_{\ell^\prime m}(r_1).
\end{eqnarray} 

Therefore, the radial wave functions obtained from Eq.~(13), corresponding to
a given $\ell$ will have an admixture of wave functions corresponding to other 
$\ell$ values of the same parity. Thus this wave function can be quite different
from that of the spherical Woods-saxon potential. However, if components of the 
admixed states of higher $\ell$ are quite weak, then the pure states of lowest 
$\ell$ can become dominant. In such a situation, one can use the solutions of 
the spherical Woods-Saxon potential corresponding to a single $\ell$ for the 
wave function $u_\ell(r_1)$ in Eq.~(11). Indeed, it has been shown in Ref.
\cite{ham04} that as the binding energies approach zero, the lowest $\ell$ 
components become dominant in the neutron orbits of the realistic deformed 
potential irrespective of the size of the deformation. In this work, we have 
made the approximation of taking $u_\ell(r_1)$ as the state of a given single 
$\ell$ value that is the solution of the Schr\"odinger equation with spherical
Woods-Saxon potential. In any case, if the spectroscopic factors of shell model 
calculations are used for a particular state, the wave functions obtained in a 
spherical basis for that state should already include the admixture of different 
$\ell$ states.  

We would like to point out here that only the structure part of the amplitude
given by Eq.~(11) is affected by the deformation in the interaction $\vv$
- the dynamical part of it remains the same as it would be in no-deformation case.
With the deformation effects introduced through Eq.~(12), analytic expressions
can be written for the structure part of the amplitude in Eq. (11). Let
us define  
\begin{eqnarray}
I_f = \int {\bf dr}_1 e^{-i{\bf Q}.{\bf r}_1} V_{bn}({\bf r}_1) u_{\ell}({r}_1) 
Y^m_{\ell}(\hat {\bf r}_1). \label{a4.8}
\end{eqnarray}
We can write (see, Ref.~\cite{shu14} for details of this derivation),
\begin{eqnarray}
I_f &=& 4\pi\sum_{l_1m_1} i^{-l_1} Y^{m_1}_{l_1}(\hat {\bf Q}) \int r^2_1 dr_1 
j_{l_1}(Qr_1)u_{\ell}(r_1)\nonumber\\
&\times&\left[V_{ws}(r_1) \delta_{l_1,\ell}\delta_{m_1,m}-\beta_{2}R V_{ws}^0
\frac{df(r_1)}{dr_1}I_1 \right]\label{a4.14},
\end{eqnarray}
where $I_1$ is defined as
\begin{eqnarray}
I_1=\int d\Omega_{r_1}Y^{0}_{2}(\hat {\bf r}_1)Y^{{m_1}\ast}_{l_1}
(\hat {\bf r}_1)Y^{m}_{\ell}(\hat {\bf r}_1)= (-1)^{m_1}\sqrt{\frac{5}{4\pi}}
\left[\frac{(2{\ell}+1)(2l_1+1)}{4\pi}\right]^{1/2}\nonumber\\
  \times \left(\begin{array}{ccc}l_1 & 2 & \ell \\ 0 & 0 & 0 \end{array}\right)
\left(\begin{array}{ccc}l_1 & 2 & \ell \\ -m_1 & 0 & m \end{array}\right), 
\label{a4.13}
\end{eqnarray}
with $|\ell-2| < l_1 < |\ell+2|$ and $m_1 = m$. Notice that there would be a
limited number of $l_1$ values to be considered, given that $\ell$ is the orbital
angular momentum of the projectile ground state. In the limit of $\beta_2 = 0$, 
the above equation would contain the first term in the square bracket [involving 
the spherical potential $V_{ws}(r_1)$] that is reduced precisely in the same form
as that obtained in Ref.~\cite{cha00} for the case where there is no deformation.

In Eq.~(14), the spherical harmonic $Y^{{m_1}\ast}_{l_1}(\hat {\bf Q})$ (where 
${\bf Q} = \gamma {\bf q_n}-\alpha{\bf K}$) can be written in terms of product 
of two spherical harmonics, one depending on $\hat {\bf q}_n$ and the other 
depending on $\hat {\bf K}$, using  Moshinsky's formula \cite{mos59}:
\begin{eqnarray}
(|{\bf Q}|)^{l_1}Y^{m_1}_{l_1}(\hat {\bf Q}) &=& \sum_{LM_L}\frac{\sqrt{4\pi}}
{\hat L}\left(\begin{array}{c} 2l_1+1\\ 2L\end{array}\right)^{1/2}
|\alpha K|^{l_1-L} (\gamma q_n)^L \nonumber\\
&\times&\left\langle l_1-L\hspace{0.1in} m_{1}-M_L\hspace{0.1in} L\hspace{0.1in} 
M_{L}|l_1\hspace{0.1in} m_{1}\right\rangle Y^{m_1-M}_{l_1-L}
(\hat {\bf K})Y^{M}_{L}(\hat {\bf q}_n),\nonumber\\
~
\end{eqnarray}
where $\left(\begin{array}{c} 2l_1+1\\ 2L\end{array}\right)$ is the binomial
coefficient and ${\hat L } = \sqrt{2L+1}$ with $L$ varying from 0 to $l_1$.
Therefore, the structure part Eq. (\ref{a4.14}), can be evaluated and would contain
the effect of the deformation of the projectile.

We once again wish to emphasize the analytic nature of our calculation at this 
point.  With the structure part given by Eq. (\ref{a4.14}), the dynamics part in
Eq.~ (11) is still given by the Bremsstrahlung integral, which can be
solved analytically.

The triple differential cross section for the reaction is related to reduced 
transition amplitude $\beta_{\ell m}$ as  
\begin{eqnarray}
\frac{d^3\sigma}{dE_{b}d\Omega_{b}d\Omega_{n}} = \frac{2\pi}{\hbar v_{a}}
\rho(E_{b},\Omega_{b},\Omega_{n})\sum_{\ell m}|\beta_{\ell m}|^2,\label{a4.1}
\end{eqnarray}
where $v_{a}$ is the $a-t$ relative velocity in the entrance channel and 
$\rho(E_{b},\Omega_{b},\Omega_{n})$ is the phase space factor appropriate to 
the three-body final state. 


\section{Results and discussions}

The formalism described in Section 2, has been employed to investigate Coulomb 
breakup of $^{37}$Mg on a Pb target at the beam energy of 244 MeV/nucleon. In 
our analysis the calculated one-neutron removal cross sections are compared with 
the corresponding data as reported in Ref.~\cite{kob14}. To calculate the Coulomb 
breakup amplitude [see, Eq.~(11)], we require the single-particle wave function 
$u_\ell(r)$ that describes the core-valence neutron relative motion in the ground 
state of the projectile (for a given neutron-core configuration). As discussed in
the previous section, we take this wave function as that of a state of a single 
$\ell$ value and obtain it by solving the Schr\"odinger equation with a spherical 
Woods-Saxon potential with radius ($r_0$) and diffuseness ($a$) parameters of 
1.24 fm and 0.62 fm, respectively. The depth of this well is adjusted to 
reproduce the valence neutron separation energy corresponding to this  state. 

\begin{table}[ht]
\begin{center}
\caption{Depth (V$_{ws}^0$) of the Woods-Saxon potential well as a function of 
$S_n$ corresponding to neutron removal from the 2$p_{3/2}$, 2$s_{1/2}$, 
1$f_{7/2}$ orbitals. The values of parameters $r_0$ and $a$ are taken to be 1.24 
fm and 0.62 fm, respectively in all the cases.}
\vspace{0.2cm}
\begin{tabular}{|c|c|c|c|}
\hline\hline
$S_n$                & V$_{ws}^0$ (2$p_{3/2}$)  & V$_{ws}^0$ (2$s_{1/2}$) & 
V$_{ws}^0$ (1$f_{7/2}$)\\ 
{\footnotesize(MeV)}     & {\footnotesize(MeV)}  & {\footnotesize(MeV)} &
\footnotesize(MeV)  \\
\hline 
0.01 & 43.97 & 24.73 & 43.36\\
0.05 & 44.18 & 25.28 & 43.43 \\
0.10 & 44.42 & 25.72 & 43.53 \\
0.15 & 44.64 & 26.06 & 43.62 \\
0.20 & 44.84 & 26.36 & 43.71 \\
0.22 & 44.92 & 26.47 & 43.75 \\
0.25 & 45.04 & 26.63 & 43.81 \\
0.30 & 45.23 & 26.88 & 43.90 \\
0.35 & 45.42 & 27.11 & 43.99 \\
0.40 & 45.60 & 27.33 & 44.09 \\
0.45 & 45.77 & 27.53 & 44.18 \\
0.50 & 45.94 & 27.73 & 44.27 \\
0.55 & 46.11 & 27.92 & 44.37 \\
0.60 & 46.28 & 28.11 & 44.46 \\
0.65 & 46.44 & 28.29 & 44.55 \\
0.70 & 46.60 & 28.46 & 44.64 \\
\hline
\hline
\end{tabular}
\end{center}
\end{table}

Various observables for the reaction have been obtained by integrating the triple 
differential cross sections [see, Eq.~(17)] over appropriate angles and energies of 
the unobserved quantity. For example, the total Coulomb one-nucleon removal cross 
section for a given $\ell j$ configuration of the valence neutron is obtained by 
integrating the triple differential cross sections over angles and energy of 
fragment $b$ and angles of the valence neutron $n$. 

The nuclei in island of inversion are expected to have significant components of 
$2p-2h$ [$\nu(sd)^{-2}(fp)^2]$ neutron intruder configurations. Indeed, in 
Ref.~\cite{kob14}, it has been argued that the valence neutron in $^{37}$Mg$_{gs}$ 
is most likely to have a spin parity ($J^\pi$) of $3/2^-$ that corresponds to the 
$2p_{3/2}$ orbital. In this work, we have considered neutron removal from the 
$2p_{3/2}$, $2s_{1/2}$ and $1f_{7/2}$ orbitals that correspond to $^{37}$Mg ground 
state $J^\pi$ of $3/2^-$, $1/2^+$, and $7/2^-$, respectively. Since the $S_n$ 
of the valence neutron in the $^{37}$Mg ground state is still uncertain, we show 
in Table~1, values of the depths of the potential well as a function of $S_n$ 
for all of the three orbitals.  
\begin{figure}[ht]
\centering
\includegraphics[height=8cm, clip,width=0.6\textwidth]{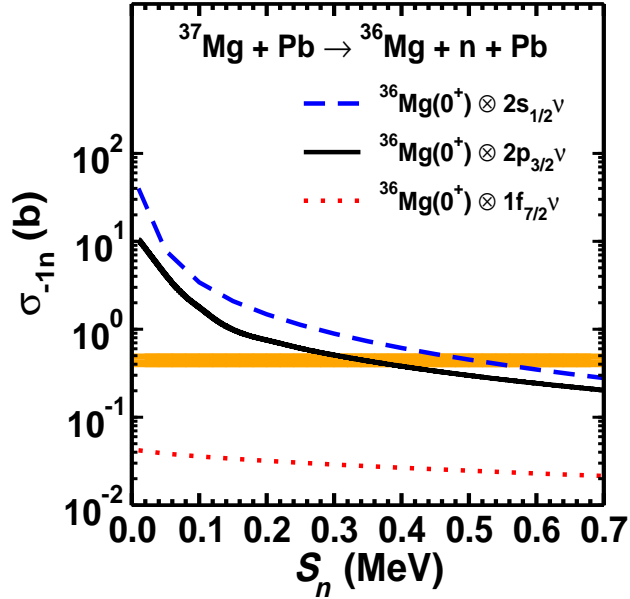}
\caption{\label{fig2} Pure Coulomb total one-neutron removal cross section, 
$\sigma_{-1n}$, in the breakup reaction of $^{37}$Mg on a Pb target at 
244 MeV/nucleon beam energy as a function of one-neutron separation energy 
$S_n$ obtained with configurations $^{36}$Mg$(0^+)\otimes 2p_{3/2}\nu$ 
(solid line),  $^{36}$Mg$(0^+)\otimes 2s_{1/2}\nu$ (dashed line) and 
$^{36}$Mg$(0^+)\otimes 1f_{7/2}\nu$ (dotted line) for $^{37}$Mg$_{gs}$ using the 
spectroscopic factors ($C^2S$) 1.0 in each case. The experimental 
cross section (taken from Ref.~\cite{kob14}) is shown by the shaded 
band.}
\end{figure}

In Fig.~2, we present the results of our calculations for the pure Coulomb 
$\sigma_{-1n}$ in the breakup reaction of $^{37}$Mg on a Pb target at the beam 
energy of 244 MeV/nucleon as a function of $S_n$ corresponding to the 
one-neutron removal from the $2p_{3/2}$, $2s_{1/2}$ and $1f_{7/2}$ orbitals. For 
$C^2S$ we have used a uniform value of one for each configuration. The shaded band 
in this figure shows the corresponding measured cross section taken from Ref.
\cite{kob14} with its width representing the experimental uncertainty. 
We note that calculated cross sections obtained with the $^{36}$Mg$(0^+)\otimes 
2p_{3/2}\nu$ and $^{36}$Mg$(0^+)\otimes 2s_{1/2}\nu$ configurations (solid and 
dashed lines, respectively in Fig.~2) overlap with the experimental band in the 
$S_n$ regions of $0.35 \pm 0.06$ MeV and $0.50 \pm 0.07$ MeV, respectively. 
Theoretical cross sections for the $2p_{1/2}$ case are almost identical to those 
of the $2p_{3/2}$ case. On the other hand, for the $^{36}$Mg$(0^+)\otimes 
1f_{7/2}\nu$ configuration there is no overlap between calculated and experimental
cross sections anywhere, which excludes the assignment of $J^\pi = 7/2^-$ to 
$^{37}$Mg$_{gs}$. Therefore, our results are consistent with the assignment of 
either of the 3/2$^-$ and 1/2$^+$ spin parity to the ground state of 
$^{37}$Mg with one-neutron separation energies in the ranges as stated above. 
The $S_n$ deduced in our work for either of these configurations is within the 
range of the evaluated value of $0.16 \pm 0.68$ MeV as reported in the most recent 
nuclear mass tabulation~\cite{wan12}.    

Nevertheless, it should be noted that there is a wide variation in the $C^2S$ 
values for these states reported in the literature. For $^{36}$Mg$(0^+)\otimes 
2p_{3/2}\nu$ and $^{36}$Mg$(0^+)\otimes 2s_{1/2}\nu$ configurations, while the 
shell model $C^2S$ values are reported to be 0.31 and 0.001, respectively
\cite{uts99}, values extracted from an analysis of the $^{37}$Mg breakup data
\cite{kob14} are 0.42$^{+0.14}_{-0.12}$ and 0.40$^{+0.16}_{-0.13}$, respectively.
In the latter work the theoretical cross sections have been computed from the 
eikonal model of Ref.~\cite{jef03} for the C target and from the (semiclassical) 
Coulomb breakup model of Ref.~\cite{nak14} for the Pb target. On the other hand, 
for the $^{36}$Mg$(0^+)\otimes 1f_{7/2}\nu$ configuration the $C^2S$ is not 
mentioned in these references. For a given neutron-core configuration $S_n$ 
extracted from the Coulomb breakup data is intimately related to the value of 
$C^2S$. Therefore, it would be interesting to investigate the $C^2S$ dependence 
of $S_n$ extracted in our study. 
\begin{figure}[ht]
\centering
\includegraphics[height=8cm, clip,width=0.7\textwidth]{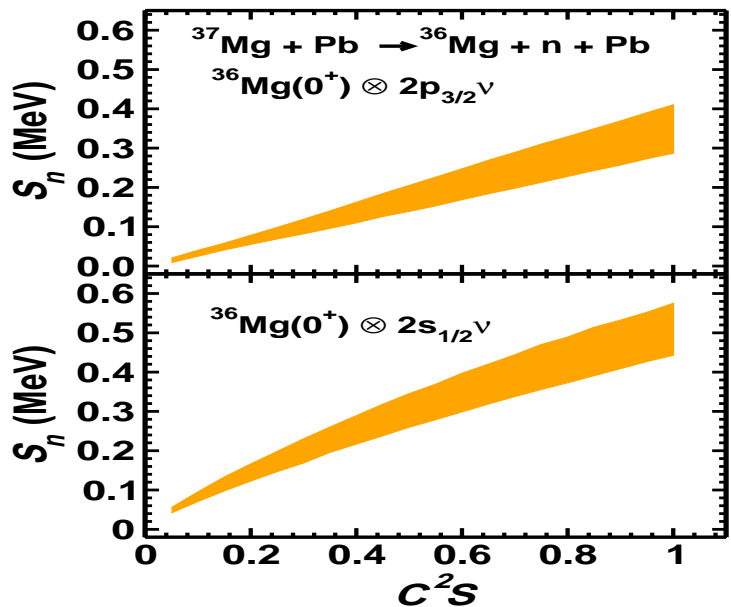}
\caption{\label{fig3} (a) $S_n$ as a function of $C^2S$ for the same reaction 
as in Fig.~2 with the $^{37}$Mg$_{gs}$ configuration of 
$^{36}$Mg$(0^+)\otimes 2p_{3/2}\nu$, (b) same as in (a) for the 
$^{36}$Mg$(0^+)\otimes 2s_{1/2}\nu$ configuration}  
\end{figure}

In Figs.~3(a) and 3(b), we show this correlation for the same reaction as in 
Fig.~2 for the configurations $^{36}$Mg$(0^+)\otimes 2p_{3/2}\nu$ and 
$^{36}$Mg$(0^+)\otimes 2s_{1/2}\nu$ of the $^{37}$Mg$_{gs}$, respectively.  In 
these calculations, for each $C^2S$ the corresponding $S_n$ is deduced from 
the region of overlap of the calculated cross section and the measured data band 
as shown in Fig.~2.  We see that $S_n$ increases steadily with increasing 
$C^2S$. Also the uncertainty in the extracted $S_n$ increases with $C^2S$, 
because at larger $C^2S$ flatter portions of the calculated cross section overlap 
with the data band that encompasses larger parts of the band. It may be mentioned  
here that in our calculations $S_n$ corresponding to the $C^2S$ of 0.42 for 
the configuration $^{36}$Mg$(0^+)\otimes 2p_{3/2}\nu$, is $0.14 \pm 0.03$ MeV, 
which is lower than the mean value of $S_n$ (0.22 MeV) obtained in Ref.
\cite{kob14} for the same $C^2S$. It is clear from this figure that for a reliable 
extraction of $S_n$ from the Coulomb breakup studies, it is essential to have 
accurate knowledge of the spectroscopic factors for different configurations. 
\begin{figure}[ht]
\centering
\includegraphics[height=8cm, clip,width=0.6\textwidth]{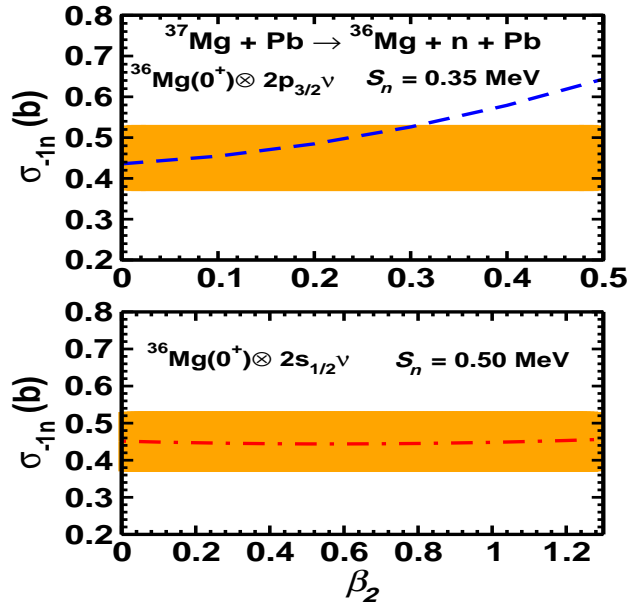}
\caption{\label{fig4} (a) $\sigma_{-1n}$ as a function of the deformation parameter
$\beta_2$ in the Coulomb breakup of $^{37}$Mg on a Pb target at the beam energy of
244 MeV/nucleon with the configuration $^{36}$Mg$(0^+)\otimes 2p_{3/2}\nu$ for 
$^{37}$Mg$_{gs}$. The $S_n$ is taken to be 0.35 MeV with $C^2S$ values being 
1.0. (b) Same as in Fig.~(a) for $^{36}$Mg$(0^+)\otimes 2s_{1/2}\nu$ configuration 
with $C^2S$ and $S_n$ of 1.0 and 0.50 MeV, respectively. In both (a) and (b) 
the experimental data (shown by the shaded region) are taken from Ref.~\cite{kob14}.}
\end{figure}
 
To substantiate the information extracted from the studies of the one-neutron 
removal cross section, it is desirable to investigate other effects and observables
to determine the most reliable configuration of $^{37}$Mg$_{gs}$. To this end, in 
Figs.~4(a) and 4(b) we investigate the effect of projectile deformation on 
$\sigma_{-1n}$ for the reaction studied in Fig.~2. As discussed earlier, the 
presence of neutrons in nearby degenerate  $j$ shells (eg., $1f_{7/2}$ and 
$2p_{3/2}$) in $^{37}$Mg that couple strongly to each other by the 
quadrupole-quadrupole interaction, can lead to the quadrupole deformation of this 
nucleus.  

In Fig.~4(a), we show our results for $\sigma_{-1n}$ as a function of $\beta_2$ 
for the $^{36}$Mg$(0^+)\otimes 2p_{3/2}\nu$ configuration of $^{37}$Mg$_{gs}$ 
with $C^2S$ values of 1.0 corresponding to a $S_n$ of 0.35, which is the mean 
value of the one-neutron separation energies extracted from the comparison of the
calculated and experimental total cross sections in Fig.~2. For $\beta_2=0$, the 
$\sigma_{-1n}$ is the same as that shown in Fig.~2 for the same value of 
$S_n$. With increasing $\beta_2$, the cross section increases, and the overlap 
between calculations and the data band ceases for $\beta_2>0.32$. Therefore, 
our calculations do not support a quadrupole deformation parameter in excess of 
0.32 for this state.

In Fig.~4(b) we show the same results for the $^{36}$Mg$(0^+)\otimes 2s_{1/2}\nu$
configuration with $C^2S$ and $S_n$ values of 1.0 and 0.50 MeV, respectively, 
The contribution of the deformation term to the cross section is substantially 
low for the $s$-wave configuration, which results in almost constant 
$\sigma_{-1n}$ as a function of $\beta_2$ as seen in Fig.~4(b). We further note 
that  in contrast to the results in Fig.~4(a), the overlap between calculated 
cross sections and the data band exists even for values of $\beta_2$ as high as 
1.2. We have checked that the situation remains the same for $\beta_2$ values 
even beyond 1.2. This result indicates that the $s$-wave configuration does not
provide any constraint on the deformation parameter $\beta_2$ in our calculations.
On the other hand, the Nilsson model study of Ref.~\cite{ham07} predicts the 
$\beta_2$ parameter of the $2s_{1/2}$ state to be below 0.3. This, however,
does not imply that the $s$-wave configuration is negated for the ground state of 
$^{37}$Mg in our calculations. It simply does not constrain the $\beta_2$ value
for this state. In any case, it is not possible to obtain more definite constraints 
on the configuration of $^{37}$Mg$_{gs}$ from a single measurement as available at 
present. 

The variation of $S_n$ with the deformation parameter $\beta_2$ is studied in
Fig.~5 for the same reaction as in Fig.~4(a) for the $^{37}$Mg$_{gs}$
configuration of $^{36}$Mg$(0^+)\otimes 2p_{3/2}\nu$ with $C^2S$ value of one. 
Several authors have argued that the deformation can lead to the enhancement of 
binding energies in the island of inversion region nuclei
\cite{cam75,war90,pov94,whi81} due to the mixing of $2\hbar \omega$ $2p-2h$ neutron 
excitations to $0\hbar \omega$ states. We notice in this figures that $S_n$ 
indeed increases with $\beta_2$. For $\beta_2 > 0.70$, the $S_n$ value exceeds
the upper limit of that evaluated in Ref.~\cite{wan12}. Therefore, for the 
$p$-wave configuration of the $^{37}$Mg$_{gs}$, the deformation parameter 
remains reasonable even for maximum predicted $S_n$. On the other hand,
with the $^{36}$Mg$(0^+)\otimes 2s_{1/2}\nu$ configuration, the $S_n$ remains
unchanged with $\beta_2$, which is obvious from Fig.~4(b). 

\begin{figure}[ht]
\centering
\includegraphics[height=8cm, clip,width=0.7\textwidth]{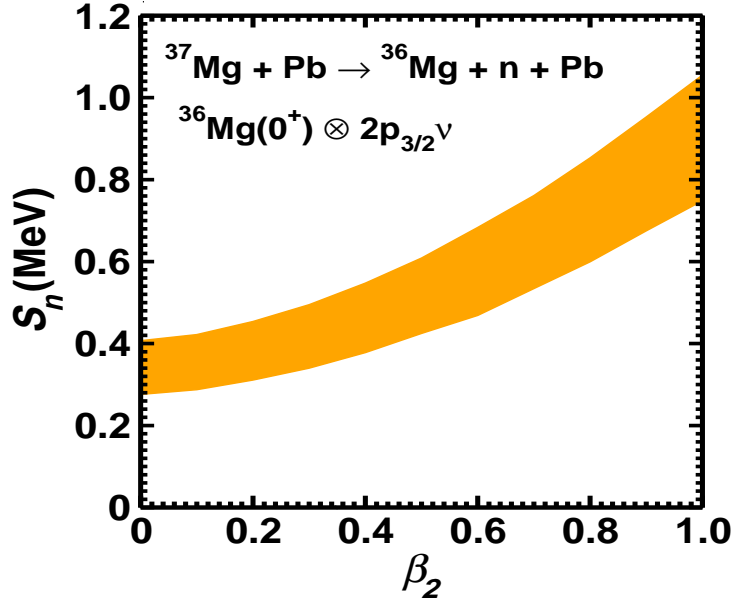}
\caption{\label{fig5} $S_n$ deduced from the comparison of our calculations 
with the experimental data as a function of the parameter $\beta_2$ for the same 
reaction as in Fig.~4, corresponding to the $^{36}$Mg$(0^+)\otimes 2p_{3/2}\nu$ 
configuration of $^{37}$Mg$_{gs}$ with $C^2S$ = 1.0. 
}
\end{figure}

\begin{figure}[ht]
\centering
\includegraphics[height=10cm, clip,width=1.0\textwidth]{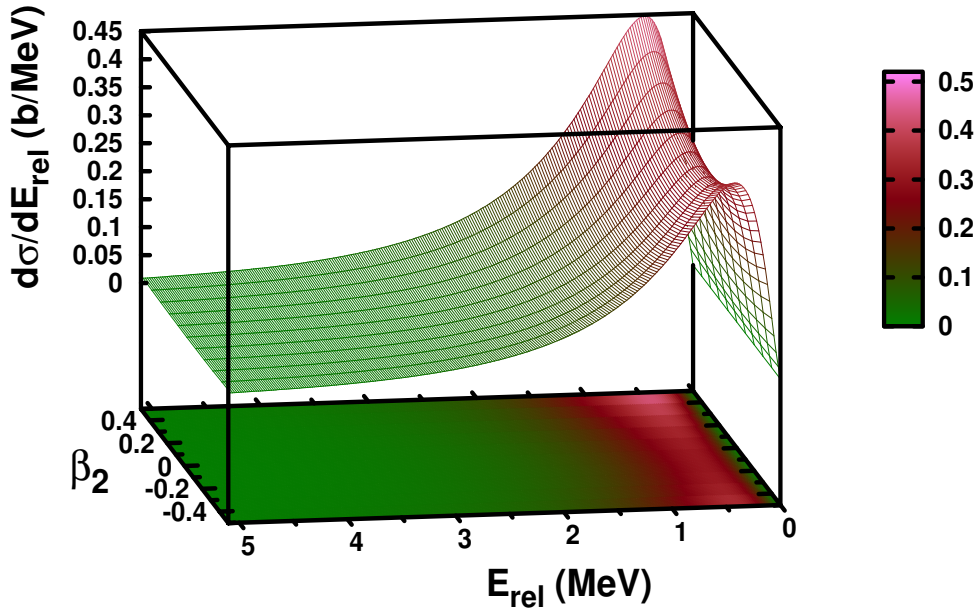}
\caption{\label{fig6} Relative energy spectra for the Coulomb breakup of $^{37}$Mg 
($J^{\pi}$ = 3/2$^-$) on a Pb target at 244 MeV/nucleon beam energy, calculated for 
different values of $\beta_2$ with C$^2S$ = 1.0 and $S_n$ = 0.35 MeV.}
\end{figure}

The investigation of more exclusive observables in the Coulomb breakup reactions of 
the projectile provides significant advantages in the understanding of its ground 
state structure. In Fig.~6, we show the $^{36}$Mg$-n$ relative energy spectra in 
Coulomb breakup of $^{37}$Mg on a Pb target at the beam energy of 244 MeV/nucleon 
as a function of the $^{36}$Mg$-n$ relative energy ($E_{rel}$) and $\beta_2$ 
simultaneously. The $^{37}$Mg$_{gs}$ configuration is $^{36}$Mg$(0^+)\otimes 
2p_{3/2}\nu$ with $C^2S$ and $S_n$ values of 1.0 and 0.35 MeV, respectively. 
We note that the height of the peak depends on the value of $\beta_2$. The position 
of the peak in this spectrum is dependent on the configuration of the projectile 
ground state, which is made more explicit in the next figure. 
\begin{figure}[ht]
\centering
\includegraphics[height=8cm, clip,width=0.7\textwidth]{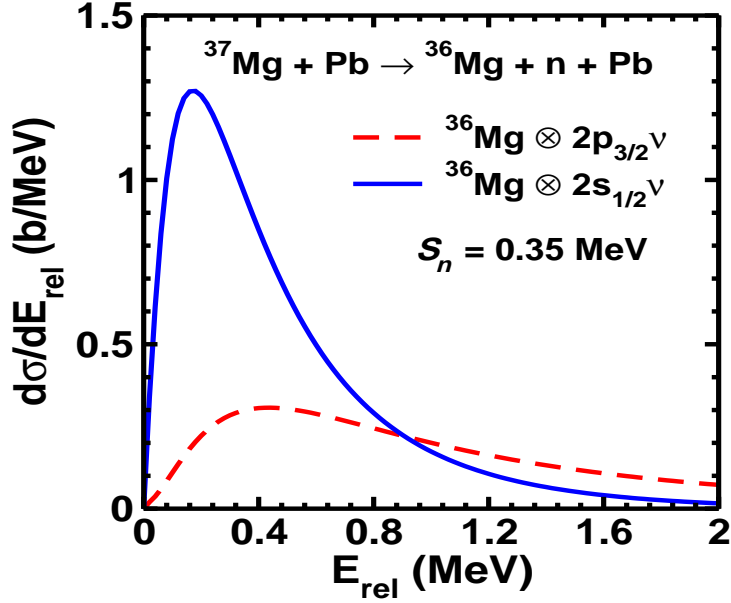}
\caption{\label{fig7} Comparison of relative energy spectra for the Coulomb 
breakup of $^{37}$Mg on Pb target at 244 MeV/nucleon beam energy corresponding 
to two different possible ground state configurations, 
$^{36}$Mg$(0^{+})$$\otimes$${2p_{3/2}}$$\nu$ (dashed line) and 
$^{36}$Mg$(0^{+})$$\otimes$${2s_{1/2}}$$\nu$ (solid line). The  values of $S_n$ 
and $C^2S$ are 0.35 MeV and 1.0, respectively for both the configurations. No 
deformation of the projectile has been included in these calculations. }
\end{figure}

In Fig.~7, we show the relative energy spectra ($d\sigma/dE_{rel}$) as a function 
of $E_{rel}$ for the same reaction as in Fig.~2 for two different configurations 
of $^{37}$Mg ground state as indicated in the figure. Since, the peak position of 
$d\sigma/dE_{rel}$ is known to depend on the value of $S_n$ 
\cite{pra99,bau03,nag05,typ05}, we have used the same values of $S_n$ and 
$C^2S$ (0.35 MeV and 1.0, respectively), for the two configurations. This ensures 
that differences seen in the relative energy differential cross sections of the 
two configurations are attributed solely to the differences in the projectile 
ground state structure. We see that the relative energy spectra obtained with two 
configurations show drastically different behavior as a function of $E_{rel}$. 
With the $s$-wave configuration, the magnitude of the cross section near the peak 
position is more than 3 times larger than that obtained with the $p$-wave one. 
Even the peak position of the two configuration are at different values of 
$E_{rel}$ - the $p$-wave cross sections peak at higher $E_{rel}$ as compared to 
those of the $s$-wave.  
\begin{figure}[ht]
\centering
\includegraphics[height=8cm, clip,width=0.7\textwidth]{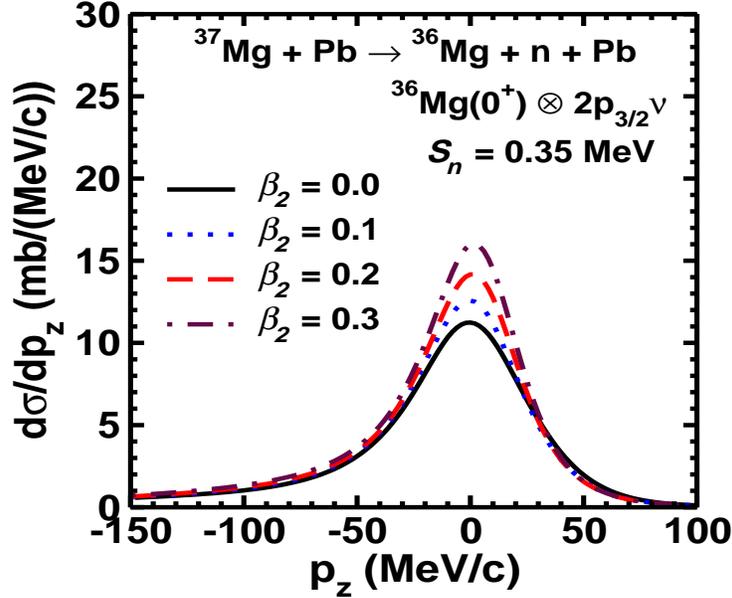}
\caption{\label{fig8} Parallel momentum distribution of $^{36}$Mg fragment in the 
Coulomb breakup of $^{37}$Mg on Pb target at 244 MeV/nucleon beam energy for 
the J$^{\pi}$ = 3/2$^-$ configuration of $^{37}$Mg$_{gs}$ with $S_n$ = 0.35 MeV 
and $C^2S$ of 1.0.} 
\end{figure}

\begin{table}[ht]
\begin{center}
\caption{Full width at half maximum of the parallel momentum distribution of 
$^{36}$Mg, obtained in Coulomb breakup of $^{37}$Mg on a Pb target at the beam 
energy of 244 MeV/nucleon. The projectile ground state corresponds to the 
configuration  of $^{36}$Mg$(0^+)\otimes 2p_{3/2}\nu$ with $S_n$ and $C^2S$
values of 0.35 MeV and 1.0, respectively.} 
\vspace{0.5cm}
\begin{tabular}{|c|c|c|}
\hline\hline
$S_n$ (MeV) & $\beta_2$ &  FWHM (MeV/c)  \\
\hline 
     & 0.0 & 54.65 \\
     & 0.1 & 50.97  \\
0.35 & 0.2 & 48.03 \\
     & 0.3 & 45.82 \\
     & 0.4 & 44.85 \\
     & 0.5 & 44.61 \\
\hline
\hline
\end{tabular}
\end{center}
\end{table}

\begin{figure}[ht]
\centering
\includegraphics[height=9cm, clip,width=0.8\textwidth]{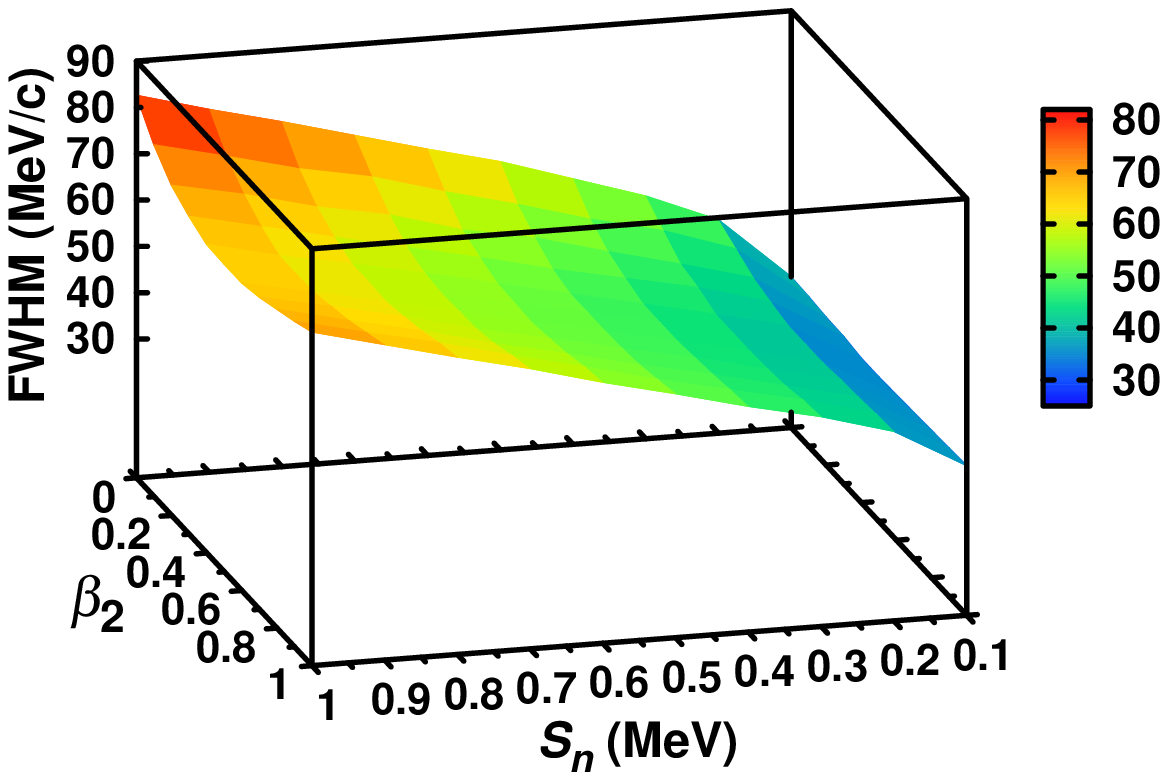}
\caption{\label{fig9} Full width at half maximum of the parallel momentum 
distribution of $^{36}$Mg, obtained in Coulomb breakup of $^{37}$Mg on a Pb target 
at the beam energy of 244 MeV/nucleon as a function of the one-neutron separation
energy $S_n$ and the quadrupole deformation parameter $\beta_2$. The projectile 
ground state corresponds to the configuration  of $^{36}$Mg$(0^+)\otimes 
2p_{3/2}\nu$ with a $C^2S$ value of 1.0.} 
\end{figure}

In view of the results shown in Fig.~7, measurements of the relative energy spectra
in the breakup reactions of $^{37}$Mg would be of great help in reducing the 
uncertainty in its ground state configuration and also in its one-neutron separation 
energy. Fixing of these quantities will lead to a better understanding of the 
quadrupole deformation of this nucleus, which also affects the height of the peak in 
the relative energy spectra.

In Fig.~8, we show the parallel momentum distribution (PMD) of the core fragment 
$^{36}$Mg in the Coulomb breakup reaction $^{37}$Mg + Pb $\to$ $^{36}$Mg + n + Pb 
at the beam energy of 244 MeV/nucleon. $^{37}$Mg$_{gs}$ was assumed to have the 
$^{36}$Mg$(0^+)\otimes 2p_{3/2}\nu$ configuration with $S_n$ and $C^2S$ 
being 0.35 MeV and 1.0, respectively. Results are shown for several values of the 
$\beta_2$ parameter. We note that the magnitude of the cross section near the 
peak position is quite sensitive to the $\beta_2$ value. Therefore, measurement
of this observable is a useful tool for putting constraints on the degree of the 
quadrupole deformation in $^{37}$Mg. 

We  note from Table 2 that the full width at half maximum (FWHM) of the PMD 
are almost the same for $\beta_2 \geq 0.30$ (~44.0 MeV/c). Even for $\beta_2$ = 
0.0, the FWHM is only about 15$\%$ larger than its value at higher $\beta_2$. 
This is very close to the FWHM of the PMD of the core fragment seen in the breakup 
reactions of the established low mass halo nuclei like $^{11}$Be and $^{19}$C.  
Therefore, the $p$-wave ($J^\pi$ = 3/2$^-$) ground state of $^{37}$Mg is highly 
likely to have a halo structure.

In Fig.~9, we show a detailed dependence of FWHM of the PMD on the 
one-neutron separation energy, $S_n$, for various values of the deformation
parameter $\beta_2$. The reactions is the same as that studied in Fig.~8 with
the same value of $C^2S$. We note that regardless of the value of $\beta_2$ 
the FWHM increases with increasing $S_n$. This is expected because with 
increasing binding energy the neutron orbits tend to become more and more like 
those of the nuclei away from the drip line. Furthermore, for most values of
$S_n$, the $\beta_2$ dependence of the FWHM of the PMD is similar to that
shown in Table 2.
 
\begin{figure}[ht]
\centering
\includegraphics[height=8cm, clip,width=0.7\textwidth]{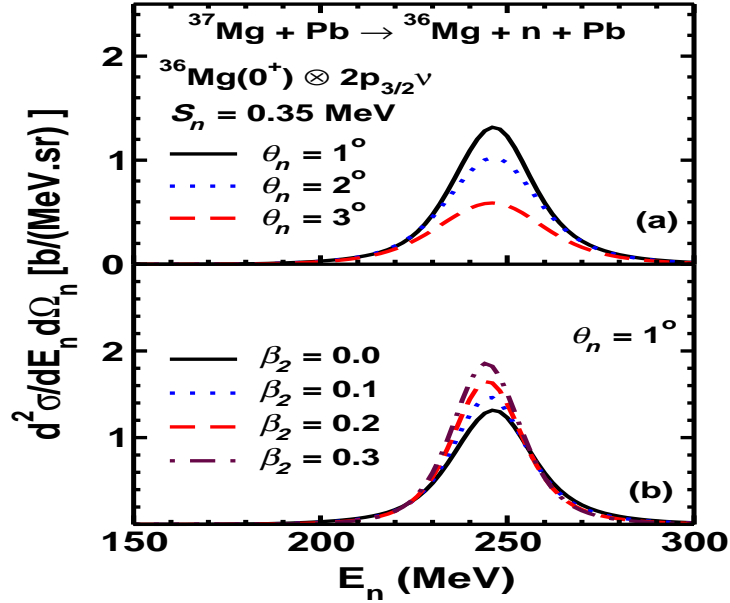}
\caption{\label{fig10} Neutron energy-angular distribution for the Coulomb breakup 
of $^{37}$Mg on a Pb target at 244 MeV/nucleon beam energy calculated for $S_n$ 
= 0.35 MeV and $C^2S$ = 1.0 for the projectile ground state configuration 
corresponding to J$^{\pi}$ = 3/2$^-$ for (a) ($\theta_n$) at 1$^\circ$ , 2$^\circ$ 
and 3$^\circ$, and (b) with different values of $\beta_2$  for $\theta_n = 1^\circ$, 
}
\end{figure}
 
In Fig.~10(a) we show the double differential cross section $d^2\sigma/dE_nd
\Omega_n$ as a function of the neutron energy for three neutron angles between 
1$^\circ-3^\circ$. The configuration of $^{37}$Mg$_{gs}$, and $C^2S$ and $S_n$ 
values were the same as those in Fig.~8. No deformation of the projectile was 
considered in these calculations (that is $\beta_2$ = 0). We see that magnitude of 
the cross section near the peak position reduces with increasing neutron angle. 
An interesting observation is that for all the three angles, the peak occurs near 
the neutron energy that correspond to the beam velocity. This is consistent with 
the picture that fragments move with the beam velocity after the breakup. If the 
charged fragment gets post-accelerated as it leaves the reaction zone (which is 
the case for the breakup reactions of stable nuclei~\cite{bau72,bau84}), one would 
expect the position of the peak in the neutron spectrum at energies below that 
corresponding to the beam velocity. We do not see this post-acceleration effect 
even if Coulomb effects have been included to all orders in the incoming and 
outgoing channels in our theory. Due to very small binding energies of the halo 
nuclei and the reactions at higher beam energies, the breakup occurs at distances 
much larger than the distance of closest approach, thus the post-acceleration 
effects are minimal~\cite{ban93,ban96,ban02}. 
 
\begin{figure}[ht]
\centering
\includegraphics[height=8cm, clip,width=0.7\textwidth]{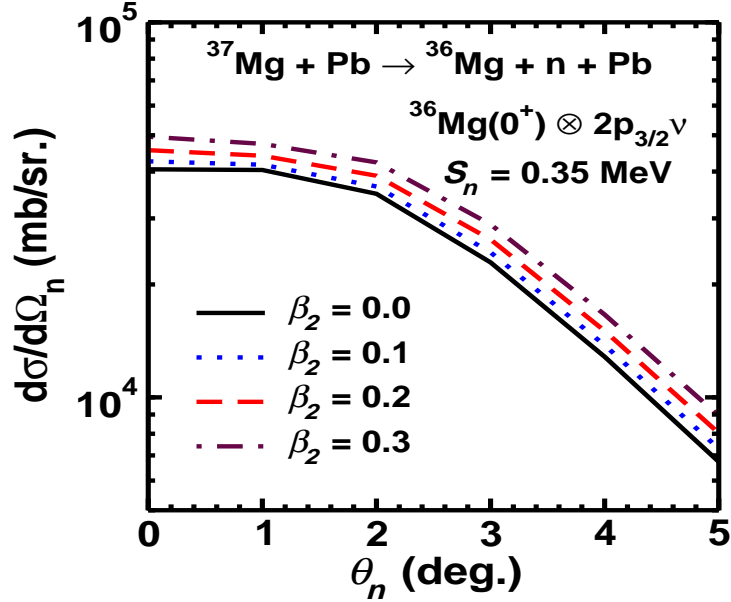}
\caption{\label{fig11} Neutron angular distribution for Coulomb breakup of 
$^{37}$Mg on a Pb target at 244 MeV/nucleon beam energy. The projectile ground 
state configuration, $S_n$ and $C^2S$ were the same as those in Fig.~10.}
\end{figure}

The effect of projectile deformation on the cross section $d^2\sigma/dE_nd\Omega_n$
is studied in Fig.~10(b) for the same reaction as in Fig.~10(a) for one angle 
of $\theta_n = 1^\circ$. It is evident from this figure that magnitude of the 
cross section increases with $\beta_2$. This is most visible near the peak position.
Therefore, measurements of the double differential cross sections are expected to 
provide additional information about the deformation of the projectile ground 
state.

The angular distributions of neutrons emitted in the projectile breakup reactions 
reflect to a great extent the momentum distribution of the fragments in the ground 
state of the projectile (see, e.g., Ref.~\cite{esb91}). Therefore, their study is 
expected to provide further information about the neutron halo structure in 
$^{37}$Mg. In Fig.~11, we show the neutron angular distribution in the Coulomb 
breakup reaction $^{37}$Mg + Pb $\to$ $^{36}$Mg + n + Pb at the beam energy of 
244 MeV/nucleon. The ground state configuration, the $S_n$ and $C^2S$ were  
the same as those in Fig.~10. The results are presented for four values of the 
$\beta_2$ parameter. We notice that cross sections drop very steeply with increasing 
neutron angle in the forward directions. The narrow angular distributions of 
neutrons below the grazing angles in the Coulomb breakup reactions of $^{37}$Mg 
reflect the small widths of the parallel momentum distribution and hence the 
large spatial extension of the valence neutron in its ground state. The effect of 
the deformation is significant at the forward angles [this was already seen in 
Fig.~10(b)]. 
 
\section{Summary and Conclusions}

In this paper we have studied the Coulomb breakup reaction $^{37}$Mg + Pb $\to$ 
$^{36}$Mg + n + Pb at the beam energy of 244 MeV/nucleon, within the framework of 
the post form finite range distorted wave Born approximation theory that is extended 
to include the projectile deformation effects. In this formalism the transition 
amplitude is factorized into two parts - one containing the dynamics of the reaction 
and the another the projectile structure informations such as the fragment-fragment 
interaction and the corresponding wave function in its ground state.  Analytic 
expressions can be written for both  parts. This formalism opens up a route to 
perform realistic quantum mechanical calculations for the breakup of neutron-drip 
line nuclei in the medium mass region that can be deformed.

We calculated the total one-neutron removal cross sections ($\sigma_{-1n}$) in 
this reaction and compared our results with the corresponding data reported in a 
recent publication~\cite{kob14} in order to determine the configuration of the 
$^{37}$Mg ground state. The analysis of this single measured cross section
already  rules out the $^{36}$Mg($0^+$) $\otimes$ ${1f_{7/2}}\nu$ configuration for 
the ground state of $^{37}$Mg. However, it does not allow to exclude either 
of the $^{36}$Mg($0^+$) $\otimes$ ${2p_{3/2}}\nu$ and $^{36}$Mg($0^+$) $\otimes$ 
${2s_{1/2}}\nu$ configurations for $^{37}$Mg$_{gs}$.  Assuming a spectroscopic 
factor of one, the extracted values of one-neutron separation energies for these 
two configurations are $0.35 \pm 0.06$ MeV and $0.50 \pm 0.07$ MeV, respectively.  
However, the deduced $S_n$ depends on the value of $C^2S$. Our study 
shows that $S_n$ rises steadily with increasing $C^2S$.  

In order to gain more insight in the ground state structure of $^{37}$Mg, we studied 
the effect of the projectile deformation on $\sigma_{-1n}$. We find that for the 
configuration $^{36}$Mg($0^+$) $\otimes$ ${2p_{3/2}}\nu$ for the $^{37}$Mg ground 
state, the calculated $\sigma_{-1n}$ overlaps with the experimental data band  
for the quadrupole deformation parameter ($\beta_2$) below 0.32. This is in line 
with the Nilsson model calculations of Ref.~\cite{ham07} where the $\beta_2$ for
this state is predicted to lie in the range 0.30 - 0.34. However, with the 
$^{36}$Mg($0^+$) $\otimes$ ${2s_{1/2}}\nu$ configuration, the overlap between 
calculations and the data occurs for even very large values of $\beta_2$. Thus 
with this configuration, our calculations are unable to put any constraint on 
deformation parameter $\beta_2$. 
 
We also calculated more exclusive observables such as the core-valence 
neutron relative energy spectra, the energy-angle and the angular distributions of 
the emitted neutron and the parallel momentum distribution of the core fragment. 
The position of the peak as well as the magnitude of the cross section near the peak
of the core-valence neutron relative energy spectra are found to be dependent on the 
configuration of the projectile ground state as well as on its deformation. Similar 
trend was also observed in the parallel momentum distribution of the core fragment.
The FWHM of this distribution are found to be of the same order of magnitude as 
those seen in the breakup of established light halo nuclei. This confirms that 
$^{37}$Mg ground state has a halo structure. The angular distribution of the 
emitted neutrons is strongly forward peaked and the cross sections in the forward 
directions, are dependent on the projectile deformation. Thus, we identified the 
observables that are more critically dependent on the ground state structure of 
the projectile. Therefore, our study is expected to provide motivation for future 
experiments on breakup reactions of the neutron rich medium mass nuclei.

In calculations of the breakup reactions of nuclei at higher beam energies, 
relativistic effects could play a role~\cite{ber05,oga09}. Our theory is essentially
non-relativistic in nature. Nevertheless, we have seen in Ref.~\cite{cha13} that 
this theory is able to reproduce well the data on the excitation energy spectra and 
the total electromagnetic one-neutron removal cross section in the breakup reaction
of $^{23}$O on a Pb target at even higher beam energy of 422 MeV/nucleon. In the
present study we did check that inclusion of relativistic effects at the kinematics 
level does not have any significant effect on the cross sections. A fully 
relativistic dynamical quantal theory of breakup reactions is still not available.
 
\noindent{\bf Acknowledgments}

This work has been supported by the Department of Science and Technology of the 
Government of India, under the grant number SR/S2/HEP-040/2012. One of the
authors (RS) is supported by the Council of Scientific and Industrial Research 
(CSIR), India. KT is supported by the Brazilian Ministry of Science, Technology 
and Innovation (MCTI-Brazil), and Conselho Nacional de Desenvolvimento Cient\'ifico 
e Tecnol\'ogico (CNPq).


\end{document}